\documentclass{aa}
\usepackage{graphicx}
\begin{document}

\title{Quark Stars as inner engines for Gamma Ray Bursts?}

\author{Rachid Ouyed \and Francesco Sannino}

\institute{Nordic Institute for Theoretical Physics, Blegdamsvej 17,
DK-2100 Copenhagen, Denmark}
\offprints{ouyed@nordita.dk}

\date{Received/Accepted}

\abstract{
A model for 
Gamma ray bursts inner engine based on quark stars
(speculated to exist in nature) is presented. 
We describe how and why these objects 
might constitute new
candidates for GRB inner engines.
At the heart of the model is the onset of
exotic phases of quark matter at the surface of such stars,
in particular the 2-flavor color superconductivity.
A novel feature of such a phase 
is the generation of  
particles which are unstable to photon decay providing
a natural mechanism for a fireball
generation; an approach which is
fundamentally different from models
where the fireball is generated during 
 collapse
or conversion of neutron star
to quark star processes.
The model is capable of reproducing 
 crucial features of Gamma ray bursts, such as the episodic
activity of the engine (multiple and random shell emission)
and the two distinct categories of the bursts
(two regimes are isolated in the model with $\sim 2$ s and $\sim 81$ s
burst total duration).
}

\maketitle

\section{Introduction}

A central problem contributing to the Gamma-ray bursts (GRBs) mystery
is the unknown nature of the engine  powering them (Kouveliotou et al. 1995;
Kulkarni et al. 1999; Piran 1999a; Piran 1999b). 
Many have been suggested but it is fair to say that we are still
far from a definite answer. Regardless of the nature
of the engine, however, it is widely accepted that
the most conventional interpretation of the observed GRBs 
result from the conversion of the kinetic energy
of ultra-relativistic particles to radiation in an
optically thin region. The particles being
accelerated by a fireball mechanism (or explosion of
radiation) taking place
near the central engine (Goodman 1986; Shemi \& Piran 1990; Paczy\'nski 1990). 

The first challenge is to conceive of circumstances that would
create a sufficiently energetic fireball.
Conversion of neutron stars to quark
stars (Olinto 1987; Cheng \& Dai 1996;
Bombaci \& Datta 2000) has been
suggested as one possibility. 
Other models also 
involve the compact object element;
such as 
black holes (Blandford \& Znajek 1977) and coalescing
neutron stars (Eichler et al. 1989; Ruffert \& Janka 1999; Janka et
al. 1999).  
We show in this work that
the plausible existence of quark stars 
combined with the onset  of 
a newly revived
state of quark matter - called color superconductivity - in 
these objects
offers a new way of tackling the GRB
puzzle (Ouyed 2002). Here we will argue that quark stars 
might constitute new 
candidates for GRB inner engines. 

Quark matter at very high density  is
expected to behave as a color superconductor (see Rajagopal \& Wilczek 2000
for a review). 
 Associated with
superconductivity is the so-called gap energy $\Delta$
inducing the quark-quark pairing and the 
 critical temperature
($T_{\rm c}$)
above which thermal fluctuations will
wash out the superconductive state.
A novel feature of such a phase is the generation of  glueball like particles
(hadrons made of gluons) which as demonstrated
in Ouyed \& Sannino (2001) immediately decay into photons. 
If  color superconductivity  sets in at the surface
of a quark star the glueball decay
becomes a natural mechanism for a fireball
generation. 

 The paper is presented as follows: In Sect. 2 we
briefly describe the concept
of color superconductivity in quark matter. 
Glueball formation and their subsequent
two-photon decay is described.
Sect. 3 deals with 
quark stars and the onset of color superconductivity at 
their surface.  
In Sect. 4, 
we explain 
how  GRBs are powered in this picture and show
that variability (multiple shell emission) is inherent to the inner engine.
We  isolate
two GRB regimes in Sect. 5 associated with 
small and massive quark stars. 
The model's features and its predictions are summarized in
Sect. 6 while a discussion and conclusion follows in Sect. 7 where the model's
assumptions and limitations are highlighted.

\section{Color Superconductivity}

While in this paper we deal mostly with the astrophysics
aspect of the model, we nevertheless
give a brief overview of color superconductivity
and the glueball-to-photon decay process which 
leads to the fireball. The
interested reader is referred to Ouyed \& Sannino
(2001) for the
underlying physics. For a recent review see
Sannino (2002)

\subsection{2-flavour color superconductivity}

\begin{figure}[t!]
\centerline{\includegraphics[width=0.5\textwidth, angle=0]{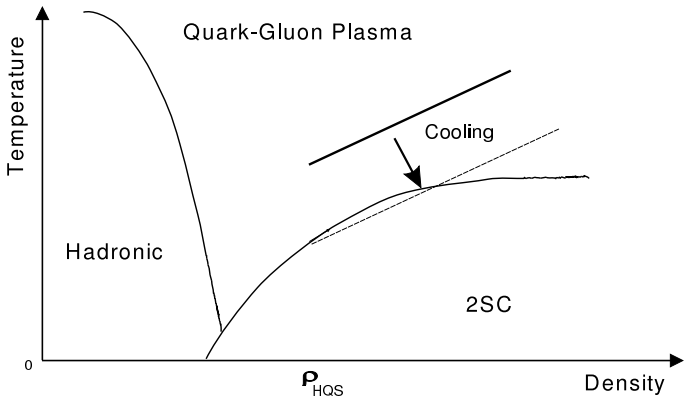}} 
\caption{A schematic representation of
a possible QCD phase diagram (Rajagopal \& Wilczek 2000).
 At high temperature and density, matter is believed to be
in a quark-gluon plasma phase (QGP).
The hadronic phase lies in the region of low
temperature and density. At very high density
but low temperature, when nuclei melt into each other,
it has been suggested that a color superconductive
phase might set in. 
2SC denotes a 2-flavor color superconductive regime.
The arrow depicts a plausible cooling path of
a HQS surface leading to the onset of color superconductivity.}
\label{Fig.1}
\end{figure}

A reasonable Quantum Chromo-Dynamics (QCD)
phase diagram (in the $\mu-T$ plane, where
$\mu$ is the chemical potential simply related
to matter density) is shown in 
Fig.~\ref{Fig.1}. At high
 temperature and density, matter is believed to be
in a quark-gluon plasma phase (QGP).
The hadronic phase lies in the region of low
temperature and density. At  high densities
but low temperatures, when nuclei melt into each other,
it is now believed that a color superconductive
phase sets in. 
This phase is characterized by the formation
of quark-quark condensate. In the 
2-flavor 
color superconductivity (2SC) the
up and down quark come into play during pairing.
Furthermore, 2SC is characterized by five out of the
eight gluons acquiring mass.   
We refer the interested reader
to Rajagopal \& Wilczek (2000)
for a review of the dynamical properties
of 2SC.

\subsection{Light GlueBalls}

The 3 massless gluons in the 2SC phase which bind into
light glueballs (LGBs) together with the
quarks up and down constitute the 2SC phase  mixture.
In Ouyed \& Sannino (2001) we
studied certain properties of these LGBs. 
Among
the properties relevant to our
present study we found,
i)  The LGBS decay into photons with an
associated lifetime of the order of $10^{-14}$ s;
ii) The mass of the LGBs is of the order of 1 MeV. 

\section{Quark stars}

We now turn to study the astrophysical consequences when such a state sets in
at the surface of quark stars. As such, we
first assume that quark stars exists in nature (further
discussed in Sect. \ref{sec6.1}) and constitutes the first major assumption in our
model.

\subsection{Hot Quark stars}

We are concerned with quark stars
born with surface temperatures above $T_{\rm c}$. 
We shall refer to these stars as ``hot'' quark stars
(HQSs) in order to avoid any confusion with strange stars
which are conjectured to exist even at zero pressure if strange matter
is the absolute ground state of strong interacting matter rather than
iron (Bodmer 1971; Witten 1984; Haensel et al. 1986; Alcock et al. 1986;
Dey et al. 1998).

We borrow the language of the MIT-bag model
formalism at low temperature
and high density to describe HQSs (Farhi \& Jaffe 1984). 
This gives a simple equation of state
\begin{equation}
P = b (\rho - \rho_{\rm HQS})c^2,
\label{one}
\end{equation}
where $b$ is a constant of model-dependent
value (close to, but generally not equal, to 1/3 of the MIT-bag
model), and   $\rho_{\rm HQS}$
is the density
at zero-pressure (the star's surface density). 
We should keep in mind that $T_{\rm c}/\mu \ll 1$
as is confirmed later.

Features of HQSs 
are - to a leading order in $T_{\rm c}/\mu$ -  identical
to that of strange stars. The latter have been studied
in details (Alcock et al. 1986;
Glendenning \& Weber 1992; Glendenning 1997).
 Of importance to our model:

i) The ``surface" of a HQS is very
different from the surface of a neutron star, or any other type of stars.
Because it is bound by the strong force, the density at the
surface  changes abruptly from zero to $\rho_{\rm HQS}$. The abrupt change 
(the thickness of the quark surface) occurs within
 about 1 fm, which is a typical strong
interaction length scale.

ii) The electrons being
bound to the quark matter by the electro-magnetic
 interaction and not by the strong
force, are able to move freely across the quark surface 
extending up to $\sim
10^3$ fm above the surface of the star.
Associated with this electron layer is a
strong electric field ($5\times 10^{17}$ V/cm)- higher than the critical
value ($1.3\times 10^{16}$ V/cm) to make the vacuum region unstable
 to spontaneously create  $(e^{+},e^{-})$ pairs.

iii) The presence of normal matter (a crust made of ions)
at the surface of the quark star is subject to the 
enormous electric dipole. The strong positive Coulomb
 barrier prevents atomic
nuclei bound in the nuclear crust from coming into
direct contact with the quark
core. The crust is suspended above the
vacuum region.

iv) One can show that the maximum
mass of the crust cannot exceed $M_{\rm crust}\simeq 5\times 10^{-5}M_{\odot}$
set by the requirement that if the density in the inner
crust is above the neutron drip
density ($\rho_{\rm drip}\simeq 4.3\times 10^{11}$ g/cc),
free neutrons will gravitate to the surface of
the HQS and be converted to quark matter.
This is due to the fact
that neutrons can easily penetrate the
Coulomb barrier and are
readily absorbed.

\subsection{Cooling and 2SC layer formation}

The HQS surface layer might enter the 2SC phase  
as illustrated in Fig.~\ref{Fig.1}.
 In the QCD phase diagram (Fig.~\ref{Fig.2}), 
 ($\rho_{{\bf B_{0}}}$, $T_{{\bf B_{0}}}$) 
is the critical point beyond
which one re-enters the QGP phase
(the extent of the 2SC layer into the star). 
The star consists of a QGP phase surrounded by
a 2SC layer where the photons (from
the LGB/photon decay) leaking from the
surface of the star 
provides the dominant cooling
source. 
This picture, as illustrated in Fig.~\ref{Fig.2}, is only valid 
if neutrino cooling in the 2SC phase is heavily
suppressed as to become slower than the
photon cooling.  
Unfortunately, 
the details of
neutrino cooling in the 2SC phase is still a topic
of debate and studies (Carter \& Reddy 2000; Schaab et al. 2000
to cite only few). One can only assume such a scenario which 
 constitutes  the second major assumption
in our model.
In Sect. \ref{cooling},
we discuss the remaining alternative when
photon cooling is dwarfed by neutrino cooling.

\subsection{LGBs decay and photon thermalization}

The photons from LGB decay are generated
at energy $E_{\gamma} < T_{\rm c}$
 and find themselves immersed in a 
degenerate quark gas. They quickly gain energy via
the inverse Compton process and become thermalized
to $T_{\rm c}$. We estimate the photon mean free path to be smaller
than few hundred Fermi (Rybicki \& Lightman 1979; Longair 1992)
 while the 2SC layer is measured in meters (see
Sect. \ref{sec4.2}). A local thermodynamic equilibrium is thus reached
with the photon luminosity given by that of  a black body radiation,
\begin{equation}
L_{\gamma} = 3.23\times 10^{52}\ {\rm ergs\ s}^{-1}\
({R_{\rm HQS}\over 5\ {\rm km}})^{2}
({T_{\rm c} \over 10\ {\rm MeV}})^4\ .
\label{three}
\end{equation}
The energy for a single 2SC event is thus
\begin{equation}
\Delta E_{\rm LGB}  = \delta_{\rm LGB} M_{\rm 2SC}c^2,
\label{four}
\end{equation}
where $M_{\rm 2SC} = \delta_{\rm 2SC} M_{\rm HQS}$ 
is the  portion of the star in 2SC. Here,
$\delta_{\rm 2SC}$ depends on the star's mass
while $\delta_{\rm LGB}$  represent
the portion of the 2SC that is in LGBs (intrinsic
property of 2SC; see Ouyed \& Sannino 2001).  The emission/cooling
 time is then 
\begin{equation}
\Delta t_{\rm cool} = {\epsilon M_{\rm HQS}c^2\over L_{\gamma}}\ ,
\label{five}
\end{equation}
with $\epsilon = \delta_{\rm 2SC}\delta_{\rm LGB}$.

\section{Powering Gamma-Ray Bursts}

\subsection{Fireball and baryon loading}

The fireball stems from the LGB
decay and photon thermalization.
The photons  are emitted from the star's surface into 
the vacuum region beneath the inner crust ($\sim 10^{3}$ fm in
size). Photon-photon interaction occurs in a much longer
time than the vacuum region crossing time. Also, the cross-section
for the creation of pairs through interactions with
the electrons in the vacuum region is negligible
(Rybicki \& Lightman 1979; Longair 1992).  The
fireball energy is thus directly deposited in the crust.  If its energy
density, $a T_{\rm c}^4$ (with $a$ being the radiation density constant),
exceeds that of the gravitational energy density
in the crust, energy outflow in the form
of ions occurs. 
One can show  that
the condition
\begin{equation}
a T_{\rm c}^4 > {G M_{\rm HQS}\over R_{\rm HQS}} \rho_{\rm crust}\ ,
\label{seven}
\end{equation}
where $\rho_{\rm crust}$ is the crust density
and $G$ the gravitational constant, is equivalent to
\begin{equation}
({T_{\rm c}\over 30\ {\rm Mev}})^4 >
({M_{\rm HQS}\over M_{\odot}})
({5\ {\rm km}\over R_{\rm HQS}})
({\rho_{\rm crust}\over \rho_{\rm drip}})\ ,
\label{eight}
\end{equation}
which is always true if $T_{\rm c} > 30$ MeV.
The  fireball is thus 
loaded with  nuclei
present in the crust.
More specifically, it is the energy transfer from
photons to electrons which drag the positively charged
nuclei in the process.
Note that
the 2SC layer is not carried out during the two-photon decay process
because of the star's high gravitational energy density:
$\rho_{\rm HQS}/ \rho_{\rm drip}>> 1$.

\begin{figure}[t!]
\centerline{\includegraphics[width=0.5\textwidth, angle=0]{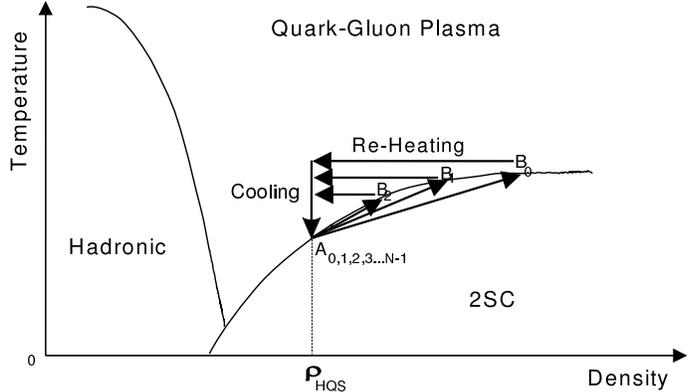}} 
\caption{
The episodic emission
as illustrated in the QCD phase diagram. 
The 2SC front spreads deep inside the star
and stops at  ${\bf B_{0}}$ before re-entering the
QGP phase. Following photon cooling,  
heat  flows from the core and
re-heats the surface. 
The star then starts cooling until
${\bf A_{1}}$ is reached at which
point  the stage is  set for the 2SC/LGB/photon
 process to start all over again
(${\bf A_{1}}\rightarrow {\bf B_{1}}$)
resulting in another emission.}
\label{Fig.2}
\end{figure}

\subsection{Episodic behavior}

The star's surface pressure is reduced following photon emission\footnote{The
pressure gradient in the 2SC layer is $\Delta p \propto
(8-5) T_{\rm c}^4$ (Farhi \& Jaffe 1984) where the
massless gluons (3 out of 8) have
been consumed by the LGB/photon process.}. 
A heat  and mass flux  is thus triggered from the QGP phase
to the 2SC layer  re-heating (above $T_{\rm c}$) and destroying
 the superconductive phase.
The entire star is now  in a QGP phase (5 gluons$\rightarrow$ 8 gluons
at the surface) and
hence the cooling process can start  again.
This corresponds to the transition
$[\rho ({\bf B}_{0}),T ({\bf B}_{0})]\rightarrow 
[\rho_{\rm HQS}, T ({\bf B}_{0})]$
 in the QCD phase diagram (thermal adjustment).
  The stage is  now set
for the 2SC/LGB/photon process to start all over again
resulting in another emission.
For the subsequent emission, however,
we expect the system to evolve to point ${\bf B}_{1}$
generally located at different densities
 and temperatures than ${\bf B}_{0}$ (see Fig.~\ref{Fig.2}).
 The cycle ends after $N$ emissions when 
$\rho({\bf B}_{N})\simeq \rho_{\rm HQS}$.

The time it takes to consume most of the star (the glue
component) by this process is
\begin{equation}
t_{\rm engine}\simeq   {M_{\rm HQS}c^2\over L_{\gamma}} \simeq 
1\ {\rm s}\  ({M_{\rm HQS}\over
M_{\odot}}) ({5\ {\rm km}\over R_{\rm HQS}})^2
({30\ {\rm MeV}\over T_{\rm c}})^4\ ,
\label{totaltime}
\end{equation}
which is representative of the engine's activity.
The above assumes quick adjustment of
the star following each event, but is not necessarily the case for the most
massive stars.

\subsection{Multiple shell emission}

The episodic behavior of the star 
together with the resulting loaded
fireball (we call shell) offers a natural mechanism
for multiple shell emission if 
$T_{c} < 30$ MeV.
Indeed from Eq. (\ref{eight}) a higher $T_{\rm c}$ value  would imply
extraction of the entire crust in a single emission
and no loading of the subsequent fireballs.
Clearly, $T_{\rm c} < 30$ MeV must be
considered if multiple ejections are to
occur\footnote{Even if $T_{\rm c}$ 
turns out to be greater than 30 MeV, in which case the
entire crust will be blown away (Eq. (\ref{eight})),
one can imagine mechanisms where crust material is replenished.
By accretion, for instance, if the HQS is part of a binary.
There are also geometrical considerations where
asymmetric emission/ejection can occur
due to the rapid rotation of Quark Stars; here
only a portion of the crust is extracted at a
time.
This aspect of the model requires better knowledge of the
conditions and environments where HQSs are formed.}.

The fraction ($f$) of the crust 
 extracted in a single event is, 
\begin{equation}
\Delta M_{\rm crust} = f M_{\rm crust}\ .
\label{nine}
\end{equation}
The  shell is accelerated
with the rest of the fireball converting most
of the radiation energy into bulk kinetic energy.
The corresponding Lorentz factor we estimate to be,
\begin{equation}
\Gamma_{\rm shell} \simeq
{\epsilon M_{\rm HQS}\over f M_{\rm crust}}\ ,
\label{ten}
\end{equation}
where we used Eq. (\ref{four}) and Eq. (\ref{nine}).
$\epsilon$ and $f$ depend on the star's mass
and characterize the two emission regimes
in our model.

\subsection{Shell-shell collision}

The Lorentz factor for the $n^{\rm th}$ shell is 
\begin{eqnarray}
\Gamma_{{\rm shell}, n} &=& {\epsilon_{n} M_{{\rm HQS},n}\over f_{n}
M_{{\rm crust},n}}\nonumber\\ 
&=& ({\epsilon_{n}\over \epsilon_{n-1}})
({f_{n-1}\over f_{n}}) {(1-\epsilon_{n-1})\over (1-f_{n-1})}
 \Gamma_{{\rm shell}, n-1}\ .
\label{nineteen}
\end{eqnarray}
The ratio
$\Gamma_{{\rm shell}, n}/\Gamma_{{\rm shell}, n-1}$ is a function
of  $\epsilon$ 
(which depends on the details
of the cooling process and the spread of the 2SC front) and $f$
(mostly related to the density in the crust).
With the two parameters varying
from one emission to another, the ratio can be
randomly greater or less than 1. As such, the shells
will have random Lorentz factors and random
energies. Faster shells
will catch up with slower ones and will
collide, converting some of their kinetic energy to internal
energy.

\section{The two regimes}

When the inner crust density is the neutron drip value, one finds
a minimum mass star of $\sim 0.015M_{\odot}$. For 
masses above this critical value, the corresponding
 crusts are thin and light.
They do not exceed few kilometers in thickness.
Matter at the density of such crusts
is a Coulomb lattice of iron and nickel
all the way from the inner edge to the surface of the
star (Baym et al. 1971). For masses below
$0.015M_{\odot}$, the crust can extend up to thousands of kilometers with
densities much below the neutron drip.
This allows us to identify two distinct emission regimes
for a given $T_{\rm c}$ ($< 30$ MeV).

\subsection{Light stars ($M_{\rm HQS} < 0.015M_{\odot}$)}

These are objects whose average density is $\sim \rho_{\rm HQS}$
($M_{\rm HQS} \simeq {4\pi\over 3}R_{\rm HQS}^{3}\rho_{\rm HQS}$).
The 2SC front extends deeper inside
the star ($\delta_{\rm 2SC}\sim 1$).
The star can be represented by a system close
to ${\bf A_{0}}$ in Fig.~\ref{Fig.2}.
Each of the few emissions (defined by $\epsilon$)
is thus capable of consuming a big portion of the star.
Furthermore, the entire crust  material
can be extracted in a few 2SC/LGB/photon cycles
($\rho_{\rm crust}/\rho_{\rm drip} \ll 1$).

Using Eq. (\ref{totaltime}), the few emissions lead to  
\begin{eqnarray}
t_{\rm tot} &\simeq& {\rm fraction}\times t_{\rm engine}\nonumber\\
&\simeq& {\rm fraction}\times 0.25\ {\rm s}\ 
({M_{\rm HQS}\over 0.01\ M_{\odot}})
({1\ {\rm km}\over R_{\rm HQS}})^2
({30\ {\rm MeV}\over T_{\rm c}})^4\ ,\nonumber\\
\label{tsmall}
\end{eqnarray}
where $t_{\rm tot}$ is representative of the
observable time which takes into account
the presence of the crust. 

\subsection{Massive stars ($M_{\rm HQS} \ge 0.015M_{\odot}$)}
\label{sec4.2}

The surface density of a massive star  being
 that of a light star ($\rho_{\rm HQS}$ given
by $P=0$ in Eq. (\ref{one})), defines
a standard unit in our model. 
In other words, the  mass of the 2SC layer
in a massive star case is
\begin{equation}
\Delta M_{\rm 2SC,m}\simeq M_{\rm 2SC,l}\ ,
\label{twelve}
\end{equation}
where ``$m$'' and ``$l$'' stand for massive and light,
respectively. It implies
\begin{equation}
{\Delta R_{\rm 2SC,m}\over R_{\rm 2SC,m}} \simeq
{1\over 3} ({R_{\rm l}\over R_{\rm m}})^3
\simeq {1\over 3} ({1\ {\rm km}\over 5\ {\rm km}})^3
\simeq 0.003\ . 
\label{thirteen}
\end{equation}
For a typical star of 5 km in radius,
we then estimate a 2SC layer of about
15 meters thick (much larger than the photon mean free path thus justifying
the local thermal equilibrium hypothesis). Equivalently,
\begin{equation}
\bar{\epsilon} = {M_{\rm l}\over M_{\rm m}} \simeq 
  ({1\ {\rm km}\over 5\ {\rm
km}})^3 \simeq 0.01\ ,
\label{fourteen}
\end{equation}
where $\bar{\epsilon}$ is the average value.
This naturally account for many events
(or $N$ fireballs). The average number
of fireballs with which an entire star is consumed is thus 
\begin{equation}
N\simeq {1\over \bar{\epsilon}}\simeq 100\ .
\label{N}
\end{equation}
Since most of the crust is at densities close to the
neutron drip value, Eq. (\ref{eight}) implies that only a tiny
part of the crust surface material (where $\rho_{\rm crust}
<< \rho_{\rm drip}$) can be extracted by each of the fireballs.
This allows for a continuous loading of the fireballs.

The total observable time in our simplified approach
is thus, 
\begin{equation}
t_{\rm tot}\simeq t_{\rm engine}
= 1\ {\rm s}\, ({M_{\rm HQS}\over M_{\odot}})
({5\ {\rm km}\over R_{\rm HQS}})^2
({30\ {\rm MeV}\over T_{\rm c}})^4\ .
\label{tmassive}
\end{equation}

We  isolated two regimes:
\\

(i) Light stars $\Rightarrow$ short  emissions.
\\

(ii) Massive stars $\Rightarrow$ long emissions.
\\

It appears, according to BATSE
(Burst and Transient Source Experiment
 detector on the COMPTON-GRO satellite), that the bursts can be
classified into two distinct categories (Kouveliotou
et al. 1993): short ($< 2$ s) bursts
 and long ($> 2$ s, typically $\sim 50$ s) bursts.
The black body behavior
($T_{\rm c}^4$) inherent to our
model puts stringent constraints on the value
of $T_{\rm c}$ which best comply with these observations. 
Using $T_{\rm c}\simeq 10$ MeV, from Eq. (\ref{tsmall})
and Eq. (\ref{tmassive})
 we obtain in the star's rest frame
\begin{equation}
t_{tot} \simeq  81\ {\rm s}\ 
 ({M_{\rm HQS}\over M_{\odot}})
({5\ {\rm km}\over R_{\rm HQS}})^2\ ,
\label{seventeen}
\end{equation}
for massive stars (suggestive of long  GRBs), and
\begin{equation}
t_{\rm tot} \simeq   2\ {\rm s}\ 
({M_{\rm HQS}\over 0.01\ M_{\odot}})
({1\ {\rm km}\over R_{\rm HQS}})^2\ ,
\label{eighteen}
\end{equation}
for light stars (suggestive of short GRBs).
There is
a clear correlation (almost one to one)
between the observed burst time and the
time at which the source ejected the specific shell
(see Figure 3 in Kobayashi et al. 1997,
for example). Note that $T_{\rm c} \simeq 10\ {\rm MeV}$ implies that
only a portion of the crust is extracted. This 
  is also consistent with our previous assumption
($T_{\rm c} < 30$ MeV) and subsequent calculations.

When $T_{\rm c}\simeq 10$ MeV, Eq. (\ref{eight}) gives
$\rho_{\rm crust}/\rho_{\rm drip}\simeq 1/16$.
For an appropriate crust density  profile 
(using the equation of state given in Baym et al. 1971), from
Eq. (\ref{nine}) we find $\bar{f}\simeq 0.01$. This implies (making
use of Eq. (\ref{fourteen}))
\begin{equation}
\Gamma_{\rm shell} = 2\times 10^{5}({\epsilon\over 0.01})
({0.01\over f})\ .
\label{twenty}
\end{equation}
For massive stars then\footnote{Note that a {\it relativistic loaded fireball}
is not necessarily achieved since $\Gamma_{\rm shell} < 100$
at times. Gaps  in the
GRBs spectra are thus expected  according to our model.}
\begin{equation}
0 < \Gamma_{\rm shell} < 2\times 10^{5}\ .
\label{twentyone}
\end{equation}

For light stars, where both
$\epsilon$ and $f$ are close to unity, 
 $\Gamma_{\rm shell}\simeq 10^{5}$.
The shells are also heavier than in the
massive stars case. We thus expect
 stronger shocks from the shell-shell
collision resulting in harder bursts.
Combined with our previous results, this is suggestive
of
\\

(i) Light stars $\Rightarrow$ short and hard bursts.
\\

(ii) Massive stars $\Rightarrow$ long and soft bursts.
\\

Eq. (\ref{seventeen}) and Eq. (\ref{eighteen}) is simply Eq. (\ref{totaltime})
rescaled  to the appropriate object size. 
We separated two regimes due to intrinsic
differences in the engine and the crust.
From the engine point of view, 
massive stars generate many more emissions when compared
to light ones, and no substantial
reduction of the engine time is expected because
of the omni-presence of the crust.
Another important difference  is 
related to the physics of the multiple re-adjustments
 following each event  which is more 
pronounced for very massive stars.
The latter among other
factors is related to $\epsilon$ which can vary from
one event to another.

\section{Features and predictions}

\subsection{GRB energies}

The maximum available
energy is when the heaviest HQS
($M_{\rm HQS,max}\simeq 2 M_{\odot}$) is entirely consumed.
That is,
\begin{equation}
E_{\rm LGB,max} \simeq 4 \times
10^{54}\ {\rm ergs}\ .
\label{twentythree}
\end{equation}
The corresponding GRB energy is 
\begin{equation}
E_{\rm GRB,max} \simeq 1.6 \times
10^{54}\ {\rm ergs}\ ,
\label{twentyfour}
\end{equation}
where we used a fiducial conversion
efficiency of 40\% (Sect. \ref{sec5.4}).

Since 
$M_{\rm HQS,min} < 0.015M_{\odot}$  we conclude that,
\begin{equation}
E_{\rm LGB,min} < 3 \times 10^{52}\ {\rm ergs}\ ,
\label{twentythree2}
\end{equation}
implying
\begin{equation}
E_{\rm GRB,min} < 1.2 \times
10^{52}\ {\rm ergs}\ .
\label{twentyfour2}
\end{equation}

\subsection{GRB total duration}

From Eq. (\ref{seventeen}) and Eq. (\ref{eighteen}) we have
\begin{equation}
t_{\rm tot} \simeq  81\ {\rm s}\ ,
\label{twentyfive}
\end{equation}
for typical massive stars, and
\begin{equation}
t_{\rm tot} \simeq   2\ {\rm s}\ ,
\label{twentysix}
\end{equation}
for typical light stars. 

Our estimate of the duration time for the massive star case
should be taken as a lower limit. As we have said,
a complete model should take into account star
re-adjustments. Nevertheless, we can still
account for a wide range in GRB duration  by
an appropriate choice of different values of the mass and radius.

\subsection{Peak duration vs energy}

The peak duration ($t_{\rm p}$) is related
to the time measured by a clock on the shell
via (Fenimore \& Ramirez-Ruiz 1999), 
\begin{equation}
t_{\rm p} \simeq {t_{\rm shell}\over 2\Gamma_{\rm shell}}= 
{1\over 2\Gamma_{\rm shell}}\times 
{\Delta X_{\rm shell}\over c_{\rm s}}\ ,
\label{twentyseven}
\end{equation}
where $c_{\rm s}$ is the speed of the shock front crossing the
shell leading to the burst. We find
\begin{equation}
t_{\rm p} 
\simeq {1\over c_{\rm s}}\times {\epsilon M_{\rm HQS}\over 4\pi
R_{\rm HQS}^2\rho_{\rm crust}}\ .
\label{twentyeight} 
\end{equation}
In the above we used Eq. (\ref{ten})
and $M_{\rm shell}=\Delta M_{\rm crust}= 4\pi R_{\rm crust}^2
\, \Delta X_{\rm crust}\, \rho_{\rm crust}$.  The
 expression on the right is $\propto \Delta t_{\rm cool}$
as can be seen from Eq. (\ref{five}) and in our model is constant for
a given star. The only parameter
which is directly linked to the shell dynamics
and energetics is $c_{\rm s}$. Shock physics gives (Ouyed \& Pudritz 1993)
\begin{equation}
c_{\rm s}^2
\propto E_{\rm int.}\ ,
\label{twentynine}
\end{equation}
where $E_{\rm int.}$ is the shell's internal energy
gained during the collision observed as the
peak's energy ($E_{\rm p}$). That is,
\begin{equation}
t_{\rm p} \propto E_{\rm p}^{-0.5}\ ,
\label{thirty}
\end{equation}
in reasonable agreement with the power law
dependence 
extracted from temporal vs energy structure in GRBs 
(index that is between $-0.37$ and $-0.46$; Fenimore et al. 1995).

\subsection{Shell dynamics}
\label{sec5.4}

Take a shell of thickness $\Delta X_{\rm crust}$ to be extracted from the
crust. The upper surface of the shell is extracted first while its lower
surface lags behind by  $(c-v_{\rm shell})t\simeq
ct/(2\Gamma_{\rm shell}^2)$ ($t$ is the time
to eject the entire shell in the star's rest frame).
 Taking into account mass conservation and the fact that  
$\Delta X_{\rm shell} = 2\Gamma_{\rm shell}^2 \Delta X_{\rm crust}$, it
is straightforward to show 
\begin{equation}
\rho_{\rm shell} \propto {1\over \Gamma_{\rm shell}^{2}}\ .
\label{twentytwo}
\end{equation}
Interestingly enough, these are the required 
conditions (including the result from Eq. (\ref{twentyone})) in the internal shock model which
lead to the highest (up to 40\%) conversion efficiency
and the most desirable temporal structure
(Kobayashi et al. 1997; Mochkovitch et al. 1995).

\section{Discussion and Conclusion}
\label{discuss}

\subsection{Existence and formation of quark stars}
\label{sec6.1}

In the last few years, thanks to the large amount of fresh
observational data collected by the new generation of X-ray and
$\gamma$-ray satellites, new observations suggest that the compact
objects associated with the X-ray pulsars, the X-ray bursters,
particularly the SAX J1808.4-3658, are good quark stars candidates (see Li
et al. 1999). While these observations/measurements 
are  hints that such objects
might  exist in nature it remains to explain their formation.
More importantly to our model, the bimodal mass distribution 
remains to be explained. 

\subsubsection{Massive stars} 

 For the massive stars the conversion of neutron stars
to quark stars is one plausible scenario (Cheng
\& Dai 1996;
Ma 1996; Bombaci \& Datta 2000).
They could also form via the direct mechanism following a
supernova collapse where
the core collapsed to a stable quark matter
instead of neutron matter (Gentile et al. 1993; Dai et al. 1995).
Both mechanisms would lead to the formation of
quark stars (strange stars to be more specific)
with masses in the solar mass range.

\subsubsection{Light stars}

The formation of small quark stars has already been
discussed (early  discussions can be found in
Alpar 1987; Glendenning et al. 1995; 
see also Chapter 10.5 in Glendenning 1997)
although these remain less understood than the massive ones.
In the case of 4U 1728-34
(where a mass of much less than $1.0M_{\odot}$ was derived; Bombaci
1999), it  seems that
accretion-induced collapse of white dwarfs is a favored 
formation mechanism. If the
quark star formed via the direct conversion mechanism then it
required too much mass (at least $\sim 0.8M_{\odot}$ to be ejected
during the conversion).

How and why stars in the 0.01$M_{\odot}$
range would form remains to be explained.
        Our arguments were solely based on theoretical
considerations related to the critical density in the
inner crust (neutron drip) as to differentiate
between small stars with thick and heavy crust versus
stars with thinner and lighter crusts.

\subsection{Neutrino cooling and HQSs}
\label{cooling}

If neutrino cooling
is shown to remain efficient in the 2SC phase
(for comparison of cooling
paths between quark stars and neutron stars
and the plausible effects of 2SC on cooling
we refer the interested reader to Schaab et al. 1997; Blaschke et al. 2000;
Blaschke et al. 2001),
we would be left with the
scenario where the entire HQS enters the 2SC phase, in which case
the 2SC/LGB/photon process
(the fireball) occurs only once and inside the entire star. 
Here, one must 
involve more complicated physics (such as that of the
crust) to account for the
episodic emissions so crucial to any model
of GRBs. It is not clear at the moment how to achieve this 
and is left as an avenue
for future research.

\subsection{2SC-II stars}

 The 2SC/LGB/photon process might proceed until
one is left with an object made entirely of 2SC. We name such
objects  {\it 2SC-II} stars\footnote{The ``II'' in 2SC 
is a simple reminder of the final state of the
star, namely the 2SC with only 5 gluons.} which are still
bound by strong interactions (their density is constant
$\sim \rho_{\rm HQS}$).
2SC-II stars carry 
an Iron/Nickel crust left over from the GRB phase. The crust
mass range is $0 <M_{\rm 2SC,crust}<5\times 10^{-5}M_{\odot}$
 depending on the efficiency of crust extraction/ejection during the GRB
phase.

BATSE  observes on average one burst per day. This corresponds,
with the simplest model - assuming no cosmic evolution of the rate - to 
about once per million years in a galaxy (Piran 1999a). 
In the Milky way we thus expect up to $10^{5}$ of 2SC-II stars.
Nevertheless, they are tiny enough ($M\le
10^{-2}M_{\odot}$, $R\le 1$ km) to be difficult to detect.

\begin{figure}[h!]
\centerline{\includegraphics[width=0.5\textwidth]{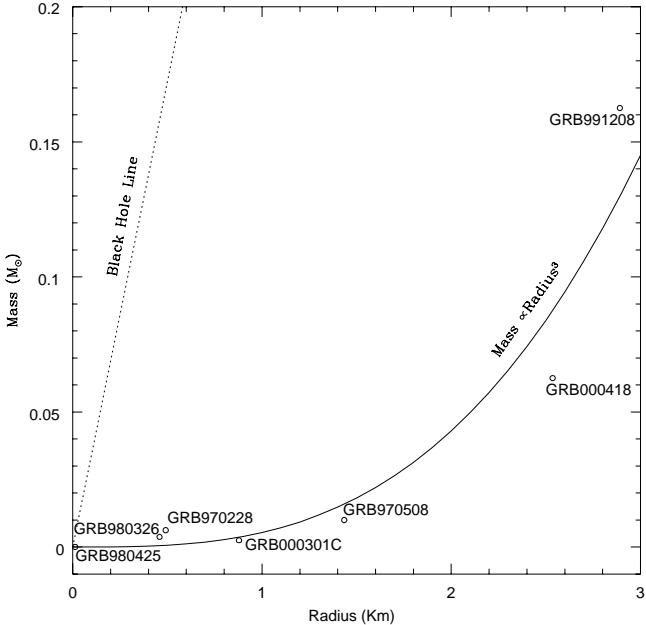}} 
\caption{The Mass$-$Radius
plane derived in our model using few existing GRBs with known
energies and total duration.  The solid curve shows the
$M_{\rm HQS}={4\pi\over 3}\rho_{\rm HQS}R_{\rm HQS}^3$ equation for $\rho_{\rm HQS} \simeq
9\rho_{\rm N}$.} 
\label{Fig.3}
\end{figure}

\subsection{The Mass$-$Radius Plane}

Take observed GRBs with
known energies and total duration.  
From the  burst total energy  $E_{\rm GRB}
\simeq 0.4 Mc^2$ we derive the mass while
 the burst total duration ($t_{\rm tot}$) 
gives us the radius (using Eq. (\ref{tmassive})
with $T_{\rm c}\simeq 10$ MeV).
In Fig.~\ref{Fig.3} we plot the resulting Mass$-$Radius.
Note that while neutron stars,
can only exists above a certain mass ($\sim 0.1M_{\odot}$;
Baym et al. 1971),
there is no lower limit to the mass of quark stars.
These would be bound by the 
strong interaction even in the absence of gravity.

The solid curve shows the
$M_{\rm HQS}={4\pi\over 3}\rho_{\rm HQS}R_{\rm HQS}^3$ equation which is a reasonable 
approximation for quark stars. 
While the GRB data set used is limited nevertheless
it seems to support the idea that extremely compact objects ($M\propto
R^3$) are behind GRBs activity within our model.

\begin{acknowledgements}
The authors thank
J. Schechter, K. Rajagopal, I. Bombaci, and F. Weber  for interesting 
and helpful discussions. 
\end{acknowledgements}

\end{document}